\documentclass[aps,prl,epsfigure,twocolumn,showpacs,superscriptaddress] {revtex4}

\usepackage{graphics}
\usepackage{amsmath}
\usepackage{amstext}
\usepackage{latexsym}
\usepackage{graphicx}
\usepackage{amsfonts}
\usepackage{amssymb}
\usepackage{epstopdf}
\usepackage{color}
\usepackage{subfigure}
\usepackage{epsfig}

\begin{document}
\title{Quantum instability and edge entanglement in the
quasi-long-range order}

\author{W. Son}
\affiliation{\text{The School of Physics and Astronomy,
University of Leeds, Leeds, LS2 9JT, United Kingdom}}
\affiliation{Centre for Quantum Technologies, National
University of Singapore, 3 Science Drive 2, Singapore 117543 }
\author{L. Amico}
\affiliation{Departemento de Física de Materiales, Universitad
Complutense de Madrid, $28040$ Madrid, Spain}
\affiliation{MATIS-INFM $\&$ Dipartimento di Metodologie Fisiche e
Chimiche (DMFCI), Universit\`{a} di Catania, viale A. Doria 6,
95125 Catania, Italy}

\author{F. Plastina}
\affiliation{\text{Dip. Fisica, Universit\`a della Calabria, \&
INFN - Gruppo collegato di Cosenza, 87036 Arcavacata di Rende
(CS) Italy}}

\author{V. Vedral}
\affiliation{\text{The School of Physics and Astronomy,
University of Leeds, Leeds, LS2 9JT, United Kingdom}}
\affiliation{Centre for Quantum Technologies, National
University of Singapore, 3 Science Drive 2, Singapore 117543 }
\affiliation{Department of Physics, National University of Singapore,
2 Science Drive 3, Singapore 117542}
\date{\today}

\begin{abstract}
We investigate the build-up of quasi-long-range order in the $XX$
chain with transverse magnetic field at finite size. As the field
is varied, the ground state of the system displays multiple level
crossings producing a sequence of entanglement jumps. Using the
partial fidelity and susceptibility, we study the transition to
the thermodynamic limit and argue that the topological order can
be described in terms of kink-antikink pairs and marked by edge
spin entanglement.
\end{abstract}
\maketitle

Strongly interacting many-body systems display a variety of zero
temperature Quantum Phase Transitions (QPT) \cite{Sachdev99}.
Quantum criticality is a property of the ground state of the
system, ultimately arising from a reshuffling of the system's
energy spectrum when control parameters are varied. In many
cases, this is accompanied by a symmetry breaking and by the
appearance of an order parameter indicating that the macroscopic
order is reached, characterized by long range correlations.
There also exist phase transitions with vanishing local order
parameter. A typical example is the
Berezinskii-Kosterlitz-Thouless (BKT) transition, occurring,
e.g. in a system of planar classical spins at a finite
temperature \cite{BKT}. The transition is characterized by a
`distortion'  of topological nature of the spatial spin
configuration, giving rise to quasi-long-range order in the
system, whose correlation functions display a power-law decay
\cite{Takahashi}. In the present paper we will be dealing with
ground states of one dimensional  systems displaying topological
order. The topological order in one dimension is peculiar and it
can be characterized by a sensitivity of the ground state by
varying the boundary conditions, giving rise to certain
solitonic edge states (see for instance \cite{AKLT}).

The prospect of practical applications, such as quantum
information processing, has led to an intense activity aimed at a
direct inspection of the properties of the ground state of a given system.
Our purpose is to characterize ground states with a non trivial
topological order through their entanglement content
\cite{Amico07}. To illustrate what we believe are generic features
of topologically ordered states with vanishing local order
parameter, we focus on a specific system: the spin $1/2$ XX chain
in external magnetic field. According to the general equivalence
between the classical-$d+1$ and quantum-$d$ criticality
\cite{Fradkin-Susskind}, this model displays a QPT from a
polarized to a BKT  phase with quasilong range order.
For such a model, we demonstrate that the quasi-long range order
manifests itself in the formation of 'entangled edge states' (with
spins at the edge of the chain sharing entanglement quite
differently from bulk spins) and that it is characterized at
finite size by an instability of the ground state determined by a
sequence of energy level crossings as the magnetic field  is
varied\cite{Buzek05}. This gives rise to sudden jumps of both the
fidelity\cite{zanardi} and the pairwise entanglement (which, thus,
behave non-analytically even at finite size). As the size of the
chain increases, the number of crossings grows until they become
dense within a sharply defined critical region. At the same time,
their effect weakens, and the behavior of an infinite chain can be
obtained through a smearing of the finite size observables, a
procedure that we apply to the partial state fidelity. Again,
however, this holds true only for bulk spins, while those at the
boundaries show a kind of rigidity and a reduced sensitivity to
the QPT.

\section{The ground state of the XX model}
We consider $N$ spin $1/2$ particles on a line, coupled by
nearest neighbor XX interaction,  with Hamiltonian
\begin{equation}
H=-J\Big[\sum_{i=1}^{N} \frac{1}{2}(\sigma_{i}^{x}\sigma_{i+1}^{x}
+\sigma_{i}^{y}\sigma_{i+1}^{y}) +B\sigma_{i}^{z}\Big]
\end{equation}
where the exchange constant $J$ has been taken as the energy unit.
In the thermodynamic limit, the system undergoes a first order
transition from a fully polarized to a critical phase with
quasi-long range order \cite{Takahashi,Katsura62}.

Assuming open boundaries (with $\sigma_{N+1}=0$), we employ the
Jordan-Wigner and Fourier transformations to introduce the fermion
operators
\begin{equation}
d_k =\sqrt{\frac{2}{N+1}} \sum_{l=1}^{N} \sin\left(\frac{\pi k
l}{N+1}\right) \, \bigotimes_{m=1}^{l-1} \sigma_m^z  \sigma_l^-,
\end{equation}
that diagonalize $H$: $H = \sum_{k=1}^{N} \Lambda_k
d_k^{\dagger}d_k + N B\openone$ with $\Lambda_k=-2B +
2\cos\left[(\pi k)/(N+1)\right]$. The $2^N$ eigenenergies and
eigenkets are $\epsilon_i \equiv\sum_{k=1}^{N} \Lambda_k
\alpha_k^{(i)} + N B$, and $|\psi_i\rangle = \Pi_{k=1}^N
(d^{\dagger}_k)^{\alpha_k^{(i)}} \, |\Omega\rangle$, with
$\alpha_k^{(i)} = \langle \psi_i |d_k^{\dagger}d_k|\psi_i\rangle
\in \{0,1\}$. The state $|\Omega\rangle$ is the fermion vacuum:
$d_k|\Omega\rangle=0 \, \forall k$.

The ground state and its energy vary with $B$, and different
ground states can be classified in terms of the number of level
crossings occurring in the system as $B$ changes. Specifically,
when $B>1$, $\Lambda_k <0$ for any $k$ and, thus, the lowest
eigenvalue is obtained by taking the state with $\alpha_k=1$
$\forall k$. The ground state energy remains $\epsilon_g^0=-NB$ as
long as $B> \cos\left[\pi/(N+1)\right]$. For these values of $B$
no level crossing occurs; the ground state is $|\psi^0_g\rangle
=\prod_{l=1}^{N} d_l^{\dagger}|\Omega\rangle$. The first crossing
occurs at $B=\cos[\pi/(N+1)]\equiv B_1$. For
$\cos[2\pi/(N+1)]<B<\cos[\pi/(N+1)]$ all of the $\Lambda_k$ are
negative except for $\Lambda_1$. Thus, the ground state energy is
obtained by {\it subtracting} its positive contribution:
$\epsilon_g^1 = \epsilon_g^0-\Lambda_1$. The corresponding
eigenstate is $|\psi^1_g\rangle= d_1 |\psi^0_g\rangle$. Letting
$B_k=\cos[k \pi/(N+1)]$ and defining the $k$-th region, $B_{k+1} <
B < B_k$, we can iterate the procedure above to find the ground
state $|\psi_g^k\rangle= d_k d_{k-1} \cdot\cdot\cdot
d_1|\psi_g^0\rangle=\prod_{l=k+1}^{N} d_l^{\dagger}|\Omega\rangle$
and its energy  as
\begin{eqnarray}
\epsilon_g^k = -(N-2k) B -2 \sum_{l=1}^{k}
\cos\left(\frac{\pi l}{N+1}\right)
\label{eq:genergy}
\end{eqnarray}
where $k$ represents the number of crossings, $1\leq k \leq N$.
For $B<B_N$, no other intersection occurs and the ground state is
simply given by the fermion vacuum state. The energy crossings are
plotted vs $B$ in Fig. \ref{fig2} (a).

To investigate the structure of the ground state and its
entanglement content, we re-write it in the spin language, using
the eigenstates of $\sigma^z_i$. For $k=0$, the ground state is
$|\varphi_g^0\rangle=|\uparrow \rangle^{\otimes N}$, which is
separable. After the 1-st crossing, it becomes
$|\varphi_g^1\rangle=\left[\sum_{l=1}^{N} S_l^1\left(
\prod_{m=1}^{l-1}\sigma^z_m\sigma_l^{-}\right)\right]|\varphi_g^0\rangle$,
where $S_l^k\equiv\sqrt{2/(N+1)}\sin\left[(\pi k l)/(N+1)\right]$.
This is an entangled state, given by a symmetric superposition of
all possible kets with one flipped spin. After $k$ level
crossings, the ground state  is given by
$|\varphi_g^k\rangle=\prod_{k'=1}^{k}\left[\sum_{l=1}^{N}
S_l^{k'}~ \left(
\prod_{m=1}^{l-1}\sigma^z_m\sigma_l^{-}\right)\right]
|\uparrow\rangle^{\otimes N}$. Explicitly
\begin{eqnarray}
\label{eq:state} |\varphi_g^k\rangle= \sum_{l_1 < l_2
<\cdot\cdot\cdot <l_k} C_{l_1 l_2 \cdot\cdot\cdot l_k} |l_1, l_2,
\cdots, l_k\rangle
\end{eqnarray}
where $|l_1, l_2, \cdots, l_k\rangle$ is the state with flipped
spins at sites $l_1,l_2,\cdots,l_k$, while the amplitudes are
given by $C_{l_1 l_2 \cdot\cdot\cdot l_k} =\sum_P (-1)^P
S_{l_1}^{P(1)} \, S_{l_2}^{P(2)}\cdots S_{l_3}^{P(3)}$, where the
sum extends over the permutation group. At each crossing point,
the ground state jumps discontinuously in the spin Hilbert space
from one symmetric subspace to another, orthogonal to the previous
one.

\begin{figure}[t]
\begin{center}
\centerline{
\includegraphics[width=3in]{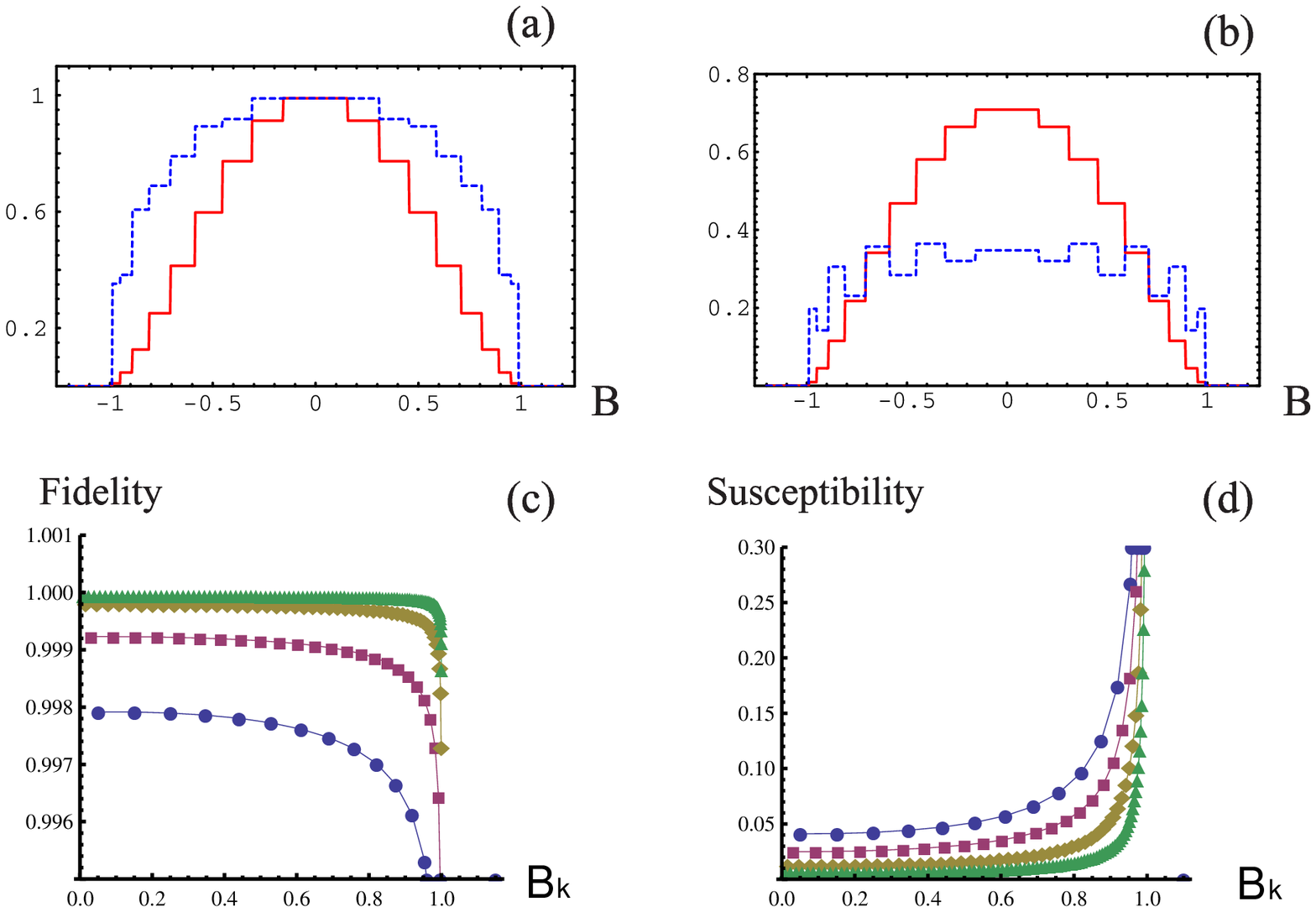}
}
\end{center}
\caption{(color online) Comparison of the entanglement shared by
spins at the beginning and at the center of the chain.(a)
One-tangle for spin at site $l$ as a function of the magnetic
field for $l=1$ (solid line) and $l=9$ (dotted). (b)
Nearest-neighbor concurrence between spins 1 and 2 (solid line),
and 9 and 10 (dotted). All of the plots are for $N=19$. (c)
Partial state fidelities and (d) Partial state fidelity
susceptibilities as a function of external magnetic field
$B_k=\cos(\frac{\pi k}{N+1})$ for  $N=30$ (blue), $N=50$ (red),
$N=100$ (yellow) and $N=200$ (green) at $l=(N+1)/2$. The
fidelity approaches 1 as $N$ increases. The parameters in the figures are
dimensionless.
} \label{fig1}
\end{figure}

\section{Finite size effects}
We now  study ground state correlation, entanglement and
fidelity at finite size. Due to the symmetry of the overall
state, the one-spin reduced density matrix is purely diagonal,
$\rho_l=\frac{1}{2}\mbox{diag}(1+\langle\sigma_l^z\rangle,
1-\langle\sigma_l^z\rangle)$; while for two spins at sites $(l,
m)$, one has $\rho_{lm}= a_+
|\uparrow\uparrow\rangle\langle\uparrow\uparrow|+ a_-
|\downarrow\downarrow\rangle\langle \downarrow\downarrow|+ b_+
|\uparrow\downarrow\rangle\langle\uparrow\downarrow|+ b_-
|\downarrow\uparrow\rangle\langle\downarrow\uparrow|+
e(|\uparrow\downarrow\rangle\langle\downarrow\uparrow|+
|\downarrow\uparrow\rangle\langle\uparrow\downarrow|)$, with
$a_{\pm} = \frac{1}{4}[1\pm\langle\sigma_l^z\rangle\pm
\langle\sigma_m^z\rangle+\langle\sigma_l^z\sigma_m^z\rangle]$, $
b_{\pm}=\frac{1}{4}[1\pm\langle\sigma_l^z\rangle\mp
\langle\sigma_m^z\rangle-\langle\sigma_l^z\sigma_m^z\rangle]$
and $ e=\frac{1}{2}\langle\sigma_l^x\sigma_m^x\rangle.$ Here,
the local magnetization and transverse correlation are given by
$\langle\sigma_l^z\rangle = 1- g_{l,l}$ and
$\langle\sigma_l^z\sigma_m^z\rangle =(1- g_{l,l})(1 -g_{m,m})-
g_{l,m}^2$, where
\begin{equation}
g_{l,m}= 2 \sum_{r=1}^{k}S^{r}_l S^{r}_m=\frac{S_l^{k+1}
S_m^k-S_l^{k} S_m^{k+1}} {2 \left[\cos \left(\frac{\pi l}{N+1}
\right)-\cos \left(\frac{\pi m}{N+1}\right) \right]},
\end{equation}
that depends on the field $B$ through the index $k$.
The longitudinal correlation function $
\langle\sigma_l^x\sigma_{m}^x\rangle$ is  \cite{Lieb61}:
\begin{equation} \langle\sigma_l^x\sigma_{m}^x\rangle =
\left |
\begin{matrix} G_{l,l+1} & G_{l,l+2} & \hdots& G_{l,m}\\
G_{l+1,l+1} & G_{l+1,l+2}  & \hdots& G_{l+1,m}  \\
\vdots & \vdots & \ddots& \\
G_{m-1,l+1} & G_{m-1,l+2} & \hdots & G_{m-1,m}
\end{matrix} \right |\; \end{equation}
where $G_{l,m}=\delta_{l,m}-g_{l,m}$. This determinant becomes of
the Toeplitz type in the thermodynamic limit. For nearest
neighbors, we simply get
$\langle\sigma_l^x\sigma_{l+1}^x\rangle=G_{l,l+1}$.

With these density matrices at hand, we discuss the entanglement
encoded in the state. Entanglement between a single spin and the
rest of the chain can be measured by the one-tangle
$\tau_l=1-\langle \sigma_l^z\rangle^2$ \cite{Wooters01}. This
quantity depends on the site for finite chain, and, as a function
of $B$, it displays jumps at each crossing point $B_k$.
Specifically: (i) $\tau_l$ equals one at zero field for every
site; (ii) the jumps near $B=\pm1$ become higher and higher moving
from the end points towards the center of the chain, see Fig.
\ref{fig1} (a). Thus, at the onset of the critical region, bulk
spins are more entangled than end ones.

To evaluate pairwise entanglement, we use the concurrence
\cite{Wooters01}, $C_{l,m}=2 \mbox{ max } \Bigl \{0, |e|-\sqrt{a_+
a_-} \Bigr \}$. Its behavior as a function of $B$ for nearest
neighboring spins with jumps at crossing points is shown in Fig.
\ref{fig1} (b). In particular, around $B=0$, the pairwise
entanglement is bigger near the end points of the chain; while the
reverse occurs at the border of the critical region.

The QPT can be further analyzed through the quantum fidelity of
ground states with slightly different fields, $B$ and $\tilde B
=B+ \delta B$. Specifically, we consider the partial state
fidelity of reduced density matrix $\rho_a(B)=\mbox{Tr}_b\rho(B)$
when the system is partitioned as $a +b$:
$F_a(B,\tilde{B})=\mbox{Tr}
\sqrt{\sqrt{\rho_a(B)}\rho_a(\tilde{B})\sqrt{\rho_a(B)}} $. $F$
characterizes the degree of change of the state as the field is
varied. For our purposes, it is sufficient to consider the
subsystem $a$ to consist of just the $l$-th
spin\cite{partial-fidelity}.
Within the critical region, $F_l$ is unit everywhere except for a
series of discrete and sharp drops at the crossing points $B=B_k$.
For large system sizes, when the crossings are dense in the
interval $|B|\leq 1$, we can perform a coarse-graining and
evaluate $F$ only at these points. As a result of this procedure,
only the drop at $B=1$ remains, while all of the intermediate ones
are smeared out, see Fig. \ref{fig1}(c). Interestingly enough,
this behavior occurs only for bulk spins. If, instead, one of the
end spins is singled out, the coarse grained fidelity stays flat.

An even more direct evidence that for large $N$ the state of the
bulk spins changes essentially in a continuous way except for the
critical point $B=1$, is obtained by looking at the fidelity
susceptibility\cite{partial-fidelity}: $ \chi_F^{l}=\lim_{\delta
B\rightarrow 0} -2 \ln F_l/(\delta B)^2,$ plotted for a bulk spin
in Fig.\ref{fig1} (b). This behavior appears related to the
essential singularity shown by the block entropy at $B=1$
\cite{block}.

\section{Infinite spin chain}
\begin{figure}[t]
{\bf (a)} \hskip3.5cm{\bf (b)}
\centerline {
\includegraphics[width=1.5in]{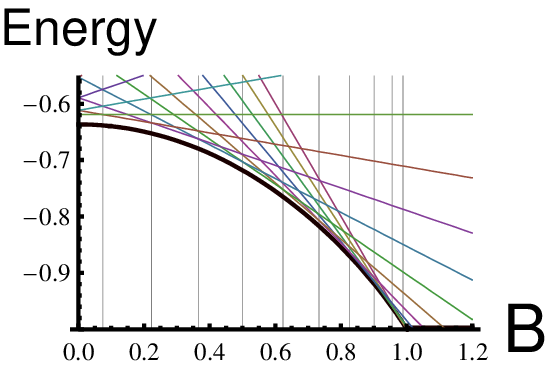}
\includegraphics[width=1.7in]{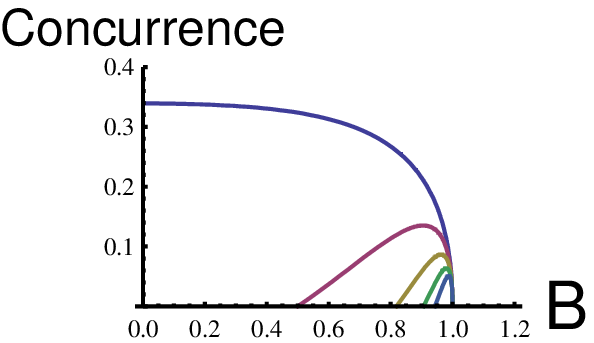}
} \caption{(color online) (a) The ground state energy per spin
(black curve) against the magnetic field $B$ in the thermodynamic
limit. The energy for $N=20$ is also plotted, together with the
intersections of the eigenvalues occurring at the crossing points
$B_k=\cos[k \pi/(N+1)]$ (vertical lines). (b) Thermodynamic limit
concurrence between two spins at distance $r=1$ (blue), $r=2$
(red), $r=3$ (yellow), $r=4$ (green) and $r=5$ (cobalt).
Entanglement between any two spins decreases as the distance
between the spins increases. It is notable that all the parameters 
in the figures are dimensionless.
}\label{fig2}
\end{figure}

For $N\rightarrow\infty$, the intervals $B_{k+1}<B<B_k$ become
infinitesimally small and $\omega\equiv (\pi k)/(N+1)$ becomes
continuous, $\omega\approx\arccos(B)$, so that the sum in
(\ref{eq:genergy}) gives an energy per spin
\begin{equation}
\displaystyle \lim_{N\rightarrow\infty}\frac{\epsilon_g(B)}{N}
=\frac{2}{\pi} \left[ B\left(\arccos B-\frac{\pi}{2}\right) -\sqrt{1-B^2} \right],
\end{equation}
which is analytic everywhere within the critical region, except
for $B=\pm 1$. From the finite size analysis, however, we know
that such region consists of dense set of crossings points (see
Fig. \ref{fig2} (a)) and therefore can be considered as a line of
continuous QPT, with the ground state driven by $B$ through
various symmetric spin subspaces with $k\approx(N+1)(\arccos
B)/\pi$ flipped spins.
The correlation functions behave differently for spins close to
the chain boundaries and for bulk spins. Setting the distance
$r=m-l$, the limit $N\rightarrow \infty$ leads to different
behaviors of $\langle \sigma^\alpha_l\sigma^\alpha_{l+r}\rangle$
depending on whether $l\gg r$  (bulk spins) or $l\ll r$ (end point
spins), \cite{MIKESKA-XX}. In particular, for $B=0$
($\omega=\pi/2$), $\langle
\sigma^z_l\sigma^z_{l+r}\rangle=(1/r^2)(l+r)^2/(l+r/2)^2$, with a
long distance behavior independent of boundary conditions. In
contrast, longitudinal correlations are sensitive to the boundary
\begin{equation}
\langle \sigma^x_l\sigma^x_{l+r}\rangle= \left \{
\begin{array}{cc}
\sqrt{2}\tilde{A}^2 r^{-1/2} & \quad l\gg r \\
4 K(\frac{l+1}{2})(\sqrt{2})\tilde{A}^2 r^{-3/4} & \quad l\ll r
\end{array} \right .
\end{equation}
where $\tilde{A}=0.6450025 $ and $K(x)$ is a function given
in\cite{MIKESKA-XX}.

For $N\rightarrow \infty$, we find $ g_{l,l+r}\simeq 2/\pi [\sin
\omega r/r -\sin \omega (2l+r)/(2l+r)],$ so that, for bulk spins
at $B=0$, $\langle \sigma^x_{l} \sigma^x_{l+r}\rangle =\left
(\frac{2}{\pi}\right )^{r}\prod_{j=1}^{[r/2]-1} \left
(\frac{4j^2}{4j^2-1}\right )^{r-2j}$ where $[x]$ is the closest
integer larger than  $x$, and $\langle \sigma^z_{l}
\sigma^z_{l+r}\rangle =\left[1-(-1)^r\right]\frac{(-2)}{\pi^2
r^2}$, which agrees with \cite{Tonegawa81}.

The concurrence $C_{l,l+r}$ can be derived from the above
correlation functions. We  find that $\lim_{l\rightarrow
\infty}C_{l,l+r}$ disappears for $r\geq 2$. Only two nearest spins
are entangled, with $\lim_{l\rightarrow \infty}C_{l,l+1}=0.339$ at
$B=0$\cite{korepin-loc}. Bulk concurrence  $\lim_{l\rightarrow
\infty}C_{l,l+r}$ between two spins at distance $1\leq r\leq 5$ is
plotted for various values of $B$ in Fig. \ref{fig2} (b), where
the decay of entanglement with the distance is also shown. For
nearest neighboring spins, the concurrence goes $\sim 1/B$,
disappearing at $B=1$. Near this point, bipartite entanglement
appears for every $r$ and any two spins in the chain become
entangled, although the magnitudes become smaller and smaller with
$r$. This is because $B=1$ is the factorizing point for the XX
model, with diverging entanglement range \cite{amicoverrucchi}.
The one-tangle $\lim_{l\rightarrow\infty}\tau_l$ behaves in a
similar way.

\section{Discussions}
In this section we show how the ground
state instability and the edge entanglement are strictly related
to the emergence of quasi long range order. It is convenient to
employ a dual basis to describe the system: $\mu_n=\prod_{m\le
n} \sigma^x_m$. Once applied to a fully polarized state, $\mu_n$
creates a topological excitation (a
kink)\cite{Fradkin-Susskind,Mikeska-dual}. Indeed, any state
with $k$ spin flips (i.e., after $k$ crossings) can be viewed as
suitable combination of  $k$ kink-antikink pairs. When $B>1$,
there are no kinks and the state is separable. Near the critical
point, with $B=1-\epsilon$, the ground state consists of a
superposition of states with a single spin-flip, or a sea of
condensed kink-antikink pairs (of infinite length for
$N\rightarrow \infty$); such a condensation gives rise to the
divergence of the concurrence range. By decreasing $B$, the size
of the kink-antikink pairs decreases and their number increases.
At $B=0$, the ground state has a single (degenerate) kink, with
half of the spins pointing down and half pointing up. The state
is `highly' symmetric and every spin is maximally entangled with
the rest of the chain, but bipartite entanglement is present
with the nearest neighbor only. This is due to the fact that the
concurrence depends (and it is always smaller than) the
longitudinal correlation function, which ultimately tends to
zero because of the presence of the kinks. Indeed, with
quasi-long range order (which arises because the spinwaves are
massless), the long range correlation function decays since the
kinks are heavy. The critical region is an instability line
because  the system is driven through different Hilbert space
sectors labelled by different quantum numbers (the eigenvalues
of the total magnetization $\sum_i\sigma^z_i$) with different
number of kinks-antikinks pairs. At finite size the phenomenon
of switching among states with a different number of kinks is
witnessed by the sequence of jumps in the entanglement, which
are smeared out for $N\rightarrow \infty$, except at $|B|=1$ as
evidenced by the partial state fidelity.

Kinks  are essentially bulk excitations and the picture above is
modified near the boundaries. Indeed, surface spins share
entanglement differently from bulk ones: for small magnetic
fields, every spin is highly entangled with the rest, but the
end-spins participate essentially to bi-partite entanglement,
while the bulk ones are rather involved in multi-partite
correlations. On the other hand, near $|B|=1$, surface spins are
less entangled than those at the center (indeed, the first jumps
of $\tau_1$ and $C_{1,2}$ are smaller than the corresponding ones
for bulk spins). Thus, end spins are less sensitive to the onset
of the critical region (as shown also by the fidelity), while they
are more entangled when the quasi long range order is fully
established. This peculiar edge entanglement is reminiscent of the
solitonic edge states found in \cite{AKLT} and could constitute a
fingerprint for topological order in one dimension, valid beyond
our specific model. Finally we note that preliminary analysis have
shown that the formalism we have developed opens the way for the
understanding the role of cluster type states \cite{Briegel01}
close to QPT, especially for quantum information purposes. Indeed,
a Hamiltonian for the cluster state can be mapped onto a model
with Ising order by means of the dual transformations employed
above. This provides a non-trivial disentangling protocol for the
cluster states \cite{Plenio07} that will be explored in a
subsequent work.

Summarizing, through the study of the entanglement content and of
the fidelity of the ground state, we have investigated the quasi
long range order in the XX model and its connection to the quantum
instability arising at finite size as a result of the presence of
topological excitations. We have also discussed the special
entanglement properties of the edge spins arguing that they are a
direct manifestation of the topological character of the QPT.

{\it Acknowledgements}- This work is supported by the National
Research Foundation \& Ministry of Education, Singapore. V. Vedral
acknowledges to EPSRC, the Royal Society and the Wolfson
Foundation for financial support. We thank J. Kwon, I. Lawrie,
H.-J. Mikeska, D. Patane' and A. Seel for discussions.


\begin{thebibliography}{99}
\bibitem{Sachdev99} S. Sachdev, Quantum Phase Transitions,
    Cambridge University Press (1999).
\bibitem{BKT} V.L. Berezinskii, Zh. Eksp. Teor. Fiz. {\bf 59},
    907 (1970) [Sov. Phys. JETP {\bf 32}, 493 (1971)]; J.M.
    Kosterlitz and D.J. Thouless, J. Phys. C {\bf 6}, 1181
    (1973).
\bibitem{Takahashi} M. Takahashi, {\it Thermodynamics of one
    dimensional solvable models}, (Cambridge Univ. press, Cambridge
    1999).
\bibitem{AKLT} I. Affleck {\it et al.}, Phys. Rev. Lett. {\bf 59},
799 (1987); Y. Kitaev, {\it cond-mat/0010440}.
\bibitem{Amico07}
    L. Amico et al.,
    \rmp {\bf 80}, 517 (2008).
\bibitem{Fradkin-Susskind} E. Fradkin and L. Susskind, Phys. Rev. D {\bf 17}, 2637
    (1978).
\bibitem{Buzek05}  A similar kind of instability has
been found for the Dicke model in  V. Bu\v{z}ek, M. Orszag and M.
Rosko, Phys. Rev. Lett. {\bf 94}, 163601 (2005).
\bibitem{zanardi}
H. T. Quan {\it et. al.},
 Phys. Rev. Lett. {\bf 96}, 140604 (2006);
P. Zanardi and N. Paunkovic, Phys. Rev. E {\bf 74}, 031123 (2006).
\bibitem{Katsura62} S. Katsura, Phys. Rev. {\bf 127}, 1508
    (1962);A. De Pasquale et al., Eur. Phys. J. Special Topics {\bf 160}, 127 (2008).
(2008), see also
\bibitem{Lieb61} E. Lieb,T. Schultz and D. Mattis, Annals of Phys.
    {\bf 16}, 407 (1961).
\bibitem{Wooters01}
W. K. Wootters, Quantum Inf. Comput. {\bf 1}, 27 (2001); W. K.
Wootters, Phys. Rev. Lett. 80, 2245 (1998).
\bibitem{partial-fidelity}
H.-Q. Zhou, R. Or\`{u}s, and G. Vidal, Phys. Rev. Lett. {\bf 100},
080601 (2008); Paunkovic et al, Phys. Rev. A {\bf 77}, 052302
(2008); H. Kwok, C. Ho and S. Gu, quant-ph/0805.3885 
;W.-L. You, Y.-W. Li, and S.-J. Gu, Phys. Rev. E {\bf 76}, 022101
(2007).
\bibitem{block} G. Vidal {\it et al.}, \prl {\bf 90}, 227902 (2003); J. I. Latorre, E.
    Rico, and A. Kitaev, Quant.Inf.Comput. {\bf 4}, 48 (2004);F. Franchini et al.,
    J.Phys. A, {\bf 40}, 8467 (2007).
\bibitem{MIKESKA-XX} H.-J. Mikeska and W. Pesch, Z. Phys. B {\bf 26}, 351 (1977).
\bibitem{Tonegawa81} T. Tonegawa, Solid State Comm. {\bf 40},
    983 (1981).
\bibitem{korepin-loc} B.-Q.Jin and V.E. Korepin,  Phys. Rev. A {\bf 69},  062314
    (2004).
\bibitem{amicoverrucchi}
L. Amico et al., Phys. Rev. A {\bf 74}, 022322 (2006).
\bibitem{Mikeska-dual} H.-J. Mikeska, S. Miyashita, and G.H. Ristow, J. Phys. C {\bf
    3}, 2985 (1991).
\bibitem{Briegel01} H. J.
    Briegel and R. Raussendorf, {\prl} {\bf 86}, 910 (2001);J. Pachos and M. Plenio,
    \prl {\bf 93}, 056402 (2004).
\bibitem{Plenio07} M. Plenio, J. Mod. Opt. {\bf 54}, 349 (2007).
\end{thebibliography}
\end{document}